\documentclass[conference]{IEEEtran}
\usepackage{epsfig}

\usepackage{ifpdf}

\usepackage{cite}

\ifCLASSINFOpdf

\else

\fi

\usepackage[cmex10]{amsmath}

\usepackage{array}

\usepackage{mdwmath}
\usepackage{mdwtab}

\usepackage{eqparbox}

\usepackage[tight,footnotesize]{subfigure}

\usepackage{fixltx2e}

\usepackage{graphicx}
\usepackage{wrapfig}

\usepackage{stfloats}

\usepackage{caption}
\usepackage{algorithm}
\usepackage{algpseudocode}
\usepackage{float}

\begin{document}
%
\title{Energy Efficient Decentralized Detection Based on Bit-optimal Multi-hop Transmission in One-dimensional Wireless Sensor Networks}

\author{\IEEEauthorblockN{Yasaman Keshtkarjahromi, Rashid Ansari, and Ashfaq Khokhar}
\IEEEauthorblockA{School of Electrical and Computer Engineering\\
University of Illinois at Chicago (UIC)\\
Chicago, Illinois 60607\\
Email: {(ykesht2, ransari, ashfaq)}@uic.edu}}


%



\maketitle

\begin{abstract}
Existing information theoretic work in decentralized detection is largely focused on parallel configuration of Wireless Sensor Networks (WSNs), where an individual hard or soft decision is computed at each sensor node and then transmitted directly to the fusion node. Such an approach is not efficient for large networks, where communication structure is likely to comprise of multiple hops. On the other hand, decentralized detection problem investigated for multi-hop networks is mainly concerned with reducing number and/or size of messages by using compression and fusion of information at intermediate nodes. In this paper an energy efficient multi-hop configuration of WSNs is proposed to solve the detection problem in large networks with two objectives: maximizing network lifetime and minimizing probability of error in the fusion node.  This optimization problem is considered under the constraint of total consumed energy.  The two objectives mentioned are achieved simultaneously in the multi-hop configuration by exploring tradeoffs between different path lengths and number of bits allocated to each node for quantization. Simulation results show significant improvement in the proposed multi-hop configuration compared with the parallel configuration in terms of energy efficiency and detection accuracy for different size networks, especially in larger networks.
\end{abstract}


%
\IEEEpeerreviewmaketitle

\section{Introduction}
Decentralized detection, cast as a hypothesis testing problem, involves making noisy observations at sensor nodes, locally quantizing these observations based on some decision rules, and then sending the quantized data to the fusion node for the final decision. For binary hypothesis, the goal is to decide between two states, $H_0$ and $H_1$ in the fusion node with minimum probability of error. Although the final decision is binary, the decisions at each sensor do not need to be binary. In other words, the quantization levels at the sensors may be more than two. Decentralized detection was first introduced by Tenney and Sandell in \cite{first}. It was extended by Tsitsiklis \cite{Tsitsiklis93}, Varshney \cite{varshney}, Viswanathan and Varshney \cite{part1} and Blum et al. \cite{Blum}. In \cite{Tsitsiklis93}, three different configurations are introduced for decentralized detection; tree, tandem, and parallel. In tandem and tree configurations, there exists a unique path from each sensor node to the fusion node and data is compressed further at each intermediate node (using the intermediate node's own observation) along the path to the fusion node. Thus, these methods would not be beneficial when the observations in the sensor nodes are independent. Also link failure in these configurations may lead to a shutdown of the whole network and failure of detection. A large body of research exists on parallel decentralized detection configuration, where each sensor node sends its quantized information directly to the fusion node. Chamberland and Veeravali, \cite{veeravalli} and \cite{decentralized}, investigate the problem of decentralized detection in sensor network applications, by considering resource constraints, such as spectral bandwidth, processing power, and cost, assuming that the information from the sensor nodes to the fusion node is transmitted over a wireless channel. In \cite{multiobjective}, the decision thresholds at each sensor node is determined to minimize the probability of error and the total consumed energy in parallel configuration. However, the major drawback of parallel configuration is the large amount of energy needed for transmitting the sensors' quantized data. In such a configuration, nodes farther away from the fusion node require more energy to send their information directly to the fusion node, thus making them less attractive for large networks. On the other hand, in multi-hop configuration, in which the quantized data can be sent to the fusion node through multiple hops, energy consumption is reduced significantly. In \cite{multihop} multi-hop configuration of decentralized detection in Wireless Sensor Networks (WSNs) is studied using fusion and compression in the intermediate nodes. However, the main aim is to reduce the number and size of messages. The number of bits used to quantize information at different nodes in the multi-hop setup is assumed to be fixed. Such assumptions limit the ability to study alternate paths and variable bit allocation to achieve lower probability of error (or higher information) at the fusion node.\\
In this paper, we study the problem of decentralized event detection, formulated as a binary hypothesis testing, for one dimensional sensor networks. In our current work, we assume that the observations at the sensors are independently distributed. We also assume that energy consumption at each sensor node is proportional to the number of quantization bits and the square of transmission distance. Thus sending data to the fusion node via multiple hops (multi-hop configuration) results in reducing the total energy consumed by the network compared with sending data directly (parallel configuration). Two optimization objectives, minimizing the probability of error in the fusion node and maximizing network lifetime, are considered under the constraint of total consumed energy to formulate our problem. The network lifetime is defined as the time it takes for the first node to deplete all its energy. Our method is based on allocating the optimal number of bits to the sensor nodes (with Maximum Likelihood Ratio (MLR) test as the decision rule at each sensor node) and determining the optimal quantization thresholds. To the best of our knowledge, no past work has investigated the benefit of optimal bit allocation amongst the sensor nodes in decentralized detection. In this paper, multi-hop as well as parallel configuration are analyzed and compared with each other. The two objectives mentioned are achieved simultaneously in multi-hop configuration by taking advantage of the two degrees of freedom provided by multi-hop configuration of WSNs, namely path selection and bit allocation.  Path selection consists of choosing the best intermediate hops to relay information to the fusion node from sensor nodes.  Bit allocation consists of allocating the optimum number of bits to each sensor node.  However in the case of parallel configuration these two objectives cannot be achieved simultaneously and thus are considered separately (with two different bit allocation methods).  This is due to the fact that in parallel configuration we are not free to choose paths from sensor nodes to the fusion node by definition.  Thus only bit allocation can be performed to satisfy either of the objectives but not both at the same time.  As shown in the simulation results, multi-hop configuration significantly outperforms the parallel configuration in terms of information quality and energy consumption. More improvements were shown for larger network sizes.\\
The rest of the paper is organized as follows. Bayesian and Neyman-Pearson formulations for decentralized detection are briefly explained in section \ref{sec:two}. In section \ref{sec:three}, the problem is formulated for parallel and multi-hop configuration and the proposed solutions are given. Simulation results are presented in section \ref{sec:five} followed by conclusion in section \ref{sec:concl}.

\section{Decentralized Detection Problem} \label{sec:two}
Let us assume $\textbf{Y} =[ Y_1, Y_2,..., Y_l,..., Y_L]$ denotes a sequence vector of measurements observed over all sensor nodes, such that $Y_j=[y_j^1,y_j^2,...,y_j^T]$ represents the measurements at sensor node $j$, $1 \le j \le L$, and $y_j^t$ denotes a single instance of measurement at sensor node $j$ at time instance $t$, $1 \le t \le T$. Let us further assume that the sequence of observations over all sensor nodes at a time instance, has the probability density function of $f_{\textbf{y}|H}(\textbf{y}|H_i), i=0,1$. Each sensor quantizes its observation according to the decision rule $\gamma _l^t: y_l^t \longrightarrow u_l^t$, and then sends the quantized information to the fusion node for final decision about the state according to the decision rule $\gamma_0: \textbf{U}=[U_1,U_2,...,U_l,...,U_L] \longrightarrow u_0$.\\
The goal in decentralized detection is to estimate the state in the fusion node with minimum probability of error. $\alpha=p(u_0=H_1|H_0)$ is the probability of error, when the actual state in the environment is $H_0$, while the decision of the fusion node is $H_1$. Similarly, $\beta=p(u_0=H_0|H_1)$ is the probability of error, when the actual state is $H_1$, while the decision of the fusion node is $H_0$. For hypothesis testing two different formulations have been used, Bayesian and Neyman-Pearson formulations. In \cite{decentralized}, the Chernoff information and Kullback-Leibler divergence are introduced as metrics to measure the probability of error in Bayesian and Neyman-Pearson formulations respectively.\\
\noindent{\bf Bayesian Formulation:} In Bayesian formulation of binary hypothesis testing, a probability is assigned to $H_0$ and $H_1$ and the goal is to minimize the probability of error, $p_e=\pi_0\alpha+\pi_1\beta$, in which $\pi_0$ is the a priori probability of state $H_0$ and $\pi_1=1-\pi_0$ is the probability of state $H_1$. The achievable upper bound for the error is given by (\cite{decentralized}):
\begin{equation}\lim_{T\rightarrow \infty} 1/T \log p_e^{(T)} \leq \log(\sum_\textbf{u} p(\textbf{u}|H_0)^s p(\textbf{u}|H_1)^{1-s}), \end{equation}
$\textbf{u}$ is the sequence of all decision rules available at the fusion node and the inequality is true for all values of $0 \leq s \leq 1$. To minimize the probability of error, we search for the decision rules that minimize the upper bound or equivalently maximize the Chernoff information at the fusion node, $C_0$ (\cite{elemntsbook}):
\begin{equation}
C_0=-\min_{0 \leq s \leq 1} [\log(\sum_\textbf{u} p(\textbf{u}|H_0)^s p(\textbf{u}|H_1)^{1-s})]
\end{equation}
\noindent{\bf Neyman-Pearson Formulation:}  In Neyman-Pearson Formulation, a constraint is imposed on one of the error probabilities, $\alpha$, and the goal is to minimize the other probability of error, $\beta$:
\begin{equation}
\underset{\gamma_1, \gamma_2, ..., \gamma_L}{\text{minimize}}
\beta (\epsilon), \;
\text{subject to}  \;
0 < \alpha < \epsilon < 1/2
\end{equation}
According to Stein's lemma (\cite{elemntsbook} and \cite{decentralized}), we have:
\begin{equation}
\lim_{\epsilon \rightarrow 0} \lim_{T\rightarrow \infty} 1/T \log \beta (\epsilon)^{(T)} = -D(p(\textbf{u}|H_0) || p(\textbf{u}|H_1)),
\end{equation}
in which $D(a||b)$ is the relative entropy of $a$ with respect to $b$ or Kullback-Leibler divergence. Therefore, for minimizing the error, we should maximize $D(p(\textbf{u}|H_0) || p(\textbf{u}|H_1))$, which is equivalent to:
\begin{equation}
D(p(\textbf{u}|H_0) || p(\textbf{u}|H_1))=\sum_\textbf{u} p(\textbf{u}|H_0) \log(\frac{p(\textbf{u}|H_0)} {p(\textbf{u}|H_1)}),
\end{equation}
\section{Proposed Multi-hop Decentralized Detection Configuration} \label{sec:three}
\begin{figure}
 \centering
 \includegraphics[scale=0.4]{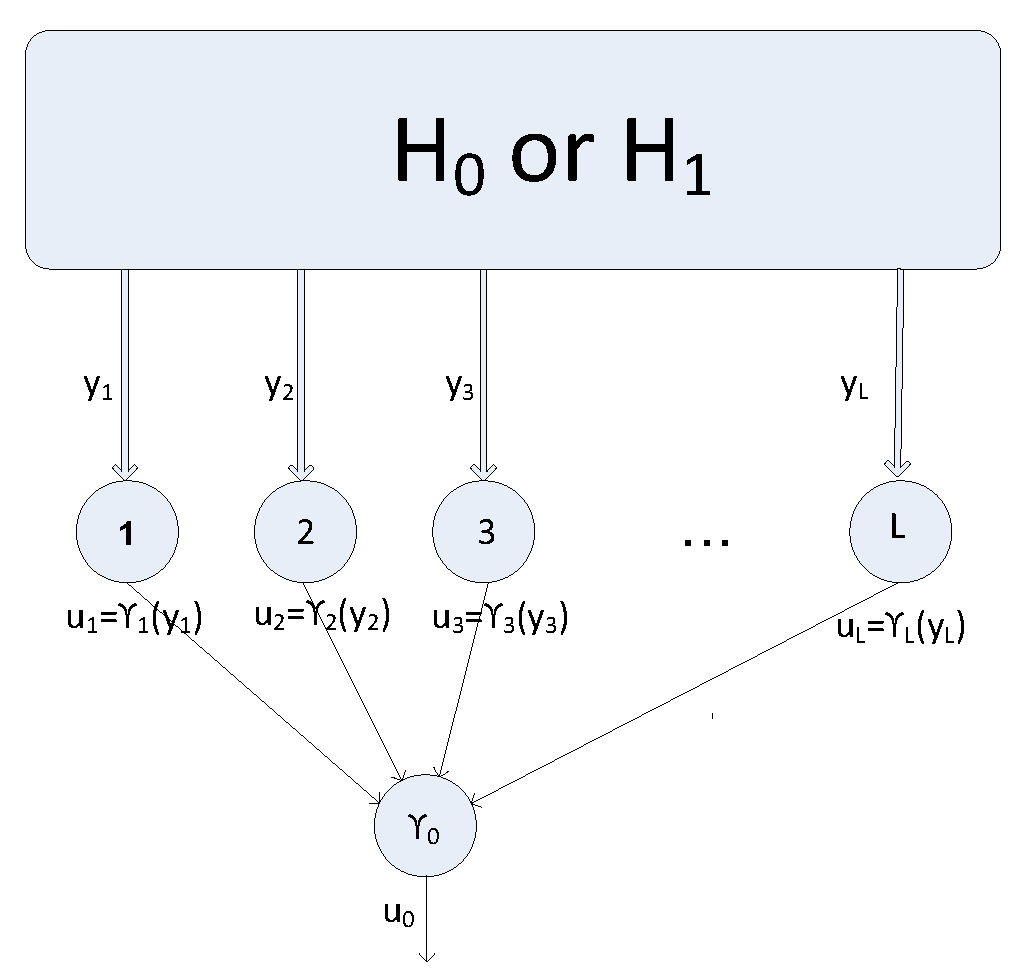}
 \caption{Parallel Decentralized Detection Configuration}
 \label{fig:Parallel}
\end{figure}

Assuming a fixed total energy budget, $E$, over all the sensor nodes, the objectives are to minimize the probability of detection error at the fusion node and maximizing the network lifetime. The first objective can be achieved by maximizing the amount of information transmitted to the fusion node and the second objective can be achieved by allocating the same amount of energy budget to each sensor node. Most of the existing research related to decentralized detection in WSN has assumed parallel configuration of nodes for communicating local decisions to the fusion node \cite{veeravalli}, \cite{decentralized}. In such a configuration, depicted in Fig. \ref{fig:Parallel}, the quantized information from the sensor nodes is sent directly to the fusion node. We assert that in parallel configuration, the two objectives cannot be simultaneously achieved. In our proposed multi-hop configuration, we maximize both the amount of information in the fusion node and the network lifetime at the same time. In addition, parallel configuration is inefficient in terms of energy and information quality compared with the proposed multi-hop configuration. The required energy for transmitting bits from one node to another is proportional to the number of transmitted bits and the square of distance between the two nodes. Therefore, we consider $bits \times distance^2$ as a metric to measure the consumed energy.
\subsection{Parallel Configuration}
In this section, the two problems, maximizing the amount of information in the fusion node and maximizing network lifetime, are formulated for parallel configuration. We can solve either of the two problems by proposing the optimal bit allocation, but not both of them at the same time.
\subsubsection{Maximizing information in the fusion node} \label{sec:infpar}
The accuracy of estimation at the fusion node depends on the amount of information the sensor nodes relay to the fusion node, which increases with an increase in the number of bits allocated to each sensor node to quantize its observation. So, both the probability of error at the fusion node and the energy consumed for transmitting the sensors' decisions to the fusion node are functions of the number of quantization bits at each sensor node. Here, we assume that the observations of the sensors are conditionally independently distributed given each state: $p(\textbf{u}|H_i) = \prod _{l=1}^L p(u_l|H_i)$ and senor node $l$ quantizes its observations with the number of allocated bits equal to $M_l$. In Bayesian formulation, the Chernoff information of the fusion node is given by (\cite{decentralized}):
\begin{equation}
\begin{split}
C_0 & = -\min_{0 \leq s \leq 1} [\log(\sum_\textbf{u} p(\textbf{u}|H_0)^s p(\textbf{u}|H_1)^{1-s})]\\
& = -\min_{0 \leq s \leq 1} (\log(\prod_{l=1}^L(\sum_{{u_l}=1}^{2^{M_l}}p({u_l}|H_0)^sp({u_l}|H_1)^{1-s})))\\
&\leq \sum_{l=1}^L (-\min_{0 \leq s \leq 1} (\log(\sum_{u_l=1}^{2^{M_l}} p({u_l}|H_0)^s p({u_l}|H_1)^{1-s}))).
\end{split}
\end{equation}
The contribution of each sensor's decision to $C_0$ is no greater than: \cite{decentralized}
\begin{equation}
C_l = -\min_{0 \leq s \leq 1} (\log(\sum_{u_l=1}^{2^{M_l}} p({u_l}|H_0)^s p({u_l}|H_1)^{1-s})),
\end{equation}
$C_l$ increases with an increase in the number of allocated bits, $M_l$. Therefore, an upper bound for it is:
\begin{subequations}
\begin{gather}
\begin{flalign}
C_l<C_l^* &= \lim_{M_l\rightarrow \infty} C_l\\
&=-\min_{0 \leq s \leq 1} (\log(\int f({y_l}|H_0)^s f({y_l}|H_1)^{1-s} dy_l)).
\end{flalign}
\end{gather}
\end{subequations}
Since $C_l, l=1,2,...,L$ is an increasing function of $M_l$ and has the limit of $C_l^*$ as $M_l \rightarrow \infty$, it is a concave function of $M_l$. Our minimization problem is aimed at determining the decision rules of the sensors that maximize the Chernoff information in the fusion node subject to the constraint of total consumed energy. The decision rule at each sensor node is completely identified by its number of allocated bits and the decision regions. To solve this optimization problem, first we find decision regions for each sensor, which maximizes its Chernoff information, $C_l$, for different values of allocated bits, $M_l$. The decision rule used at each sensor node for quantization is Maximum Likelihood Ratio (MLR) test; for a fixed number of bits, $M_l$, the optimum decision regions are obtained accordingly: $\gamma_l^{\text{opt}}(M_l)=\operatorname*{arg\,max}_{\gamma_l(M_l)} C_l(M_l),$ where,
\begin{equation}
\begin{split}
C_l(M_l)=-\min_{0 \leq s \leq 1} (\log(\sum_{u_l=1}^{2^{M_l}} & (\int_{y_l \in \gamma^{-1}_l(u_l)} f({y_l}|H_0)dy_l)^s\\
 & (\int_{y_l \in \gamma^{-1}_l(u_l)} f({y_l}|H_0)dy_l)^{1-s})),
\end{split}
\end{equation}
Therefore, based on the probability density function of observations in the sensor nodes, the increasing and concave function of $C_l(M_l)$ can be obtained.\\
In the next step, we solve the bit allocation problem amongst the sensor nodes, which determines $\textbf{M}=[M_1, M_2,..., M_l, ..., M_L]$. In parallel configuration with equal bit allocation amongst nodes, data generated by the nodes farther from the fusion node consume more energy than the closer nodes to reach the fusion node. In our proposed method, bit allocation is performed based on the contribution of each quantized data to the information in the fusion node and its distance from the fusion node. Thus, the energy required for data to reach the fusion node is proportional to its contribution to the information in the fusion node. The energy required by $M_l$ bits (quantization bits from sensor node $l$'s observations with the distance of $d_l$ from the fusion node) to reach the fusion node is equal to $E_l=M_l \times d_l^2 (bits \times distance^2)$ and its contribution to the information in the fusion node equals $C_l(M_l)$. Thus, for the two sensor nodes $l_1$ and $l_2$ we have:
\begin{equation}
\frac{M_{l_1} \times d_{l_1}^2}{M_{l_2} \times d_{l_2}^2}=\frac{C_{l_1}(M_{l_1})}{C_{l_2}(M_{l_2})} \Rightarrow \frac{C_{l_1}(M_{l_1})/M_{l_1}}{C_{l_2}(M_{l_2})/M_{l_2}}=\frac{d_{l_1}^2}{d_{l_2}^2} \label{eq:bitres}\\
\end{equation}
\begin{algorithm}[!t]
\caption{Bit Allocation in Parallel Configuration for Maximizing Information in the Fusion Node}
\label{alg:parallel}
  \begin{algorithmic}[1]
	\State Number the sensor nodes in the order of their distances from the fusion node ($d_1 \geq d_2 \geq ... d_L$). \label{alg:initial}
	\State $M_1 \gets 1$. \label{part:first}
	\State $M_l \gets \operatorname*{arg\,min}_{M_l} (\frac{d_l^2}{d_1^2} \times {C_l(M_1)/M_1} - C_l(M_l)/M_l)$ for $l=2,3,...,L$, and $M_l=0,1,2,...$ (equation \ref{eq:bitres}). \label{alg:bitall}
	\While{$\sum_{l=1}^L{M_l \times d_l^2} < E$}
	\State $M_1 \gets M_1+1$.
	\State $M_l \gets \operatorname*{arg\,min}_{M_l} (\frac{d_l^2}{d_1^2} \times {C_l(M_1)/M_1} - C_l(M_l)/M_l)$ for $l=2,3,...,L$
	\EndWhile
  \If{$\sum_{l=1}^L{M_l \times d_l^2} > E$}
	 \State $M_1 \gets M_1-1$.
	\If{$M_1 \equiv 0$}
	\State Report the bit allocated to the first node as zero. 
	\State Consider the original sensor network excluding sensor node $1$; go to step \ref{alg:initial}
	\EndIf
	\EndIf
	\State Report allocated bits as obtained in the algorithm. \label{alg:finish}
  \end{algorithmic}
\end{algorithm}
In an unrealistic case, where $M_l$ can be any real value, equation \ref{eq:bitres} is an exact equality. However, in practice $M_l$ can only be an integer value. Thus, equation \ref{eq:bitres} will become an approximation. Since, $C_l(M_l)$ is a concave function, $C_l(M_l)/M_l$ decreases with an increase in $M_l$. If the observations in the sensor nodes are identically distributed, then $C_{l_1}(M)=C_{l_2}(M)=C_l(M)$. In this case, using equation \ref{eq:bitres} results in allocating more bits to the nodes with smaller $d_l$. In other words, the nodes farther from the fusion node are allotted less number of bits and nodes closer to the fusion node are allotted more bits. In our proposed method, we first determine the maximum possible value for $M_1$, the number of bits allocated to the farthest sensor node from the fusion node. Then the allocated bits to the other sensor nodes can be obtained from the ratio of their distances from the fusion node to $d_1$ (equation \ref{eq:bitres}).\\
Since $M_l$ can take only integer values, algorithm \ref{alg:parallel} is proposed for determining bit allocation amongst the sensor nodes. This algorithm, which is based on equation \ref{eq:bitres}, outputs the nearest integer values for the allocated bits that satisfy the inequality of $\sum_{l=1}^L M_l \times d_l^2 \leq E$.
\subsubsection{Maximizing network lifetime} \label{sec:lifetime}
In a parallel configuration, the nodes farther from the fusion node consume more energy than the closer nodes and therefore run out of battery sooner. Therefore, they cannot send their information to the fusion node, which results in decreased amount of information and thus increased probability of error in the fusion node. Therefore, we define the network lifetime as the time elapsed until the first sensor node in the network depletes its energy. For maximizing the network lifetime, we try to distribute the energy consumption evenly among the sensor nodes, so that each sensor node consumes the same amount of energy as the others. Therefore, from the total energy budget, $E$, the amount of energy designated to each sensor node is $E/L$, with $L$ be the number of sensor nodes in the network. Thus, in order to guarantee that the total consumed energy is less than $E$ ($\sum_{l=1}^L M_l \times d_l^2 \leq E$), we have the following solution for bit allocation ($M_l, l=1,2,...$ must be integer values) among the sensor nodes: 
\begin{equation}
M_l = \left\lfloor \frac{E/L}{d_l^2} \right\rfloor, l=1,2,...,L
\end{equation}
In practice, the number of quantization bits cannot be more than a fixed value (e.g. $8$ bits). Here, we assume that the sensor nodes are located in the network such that the allocated bits to each sensor node are not more than a fixed number of bits.
\subsection{Multi-hop Configuration}
\begin{figure}[!t]
\centering
\includegraphics[scale=0.4]{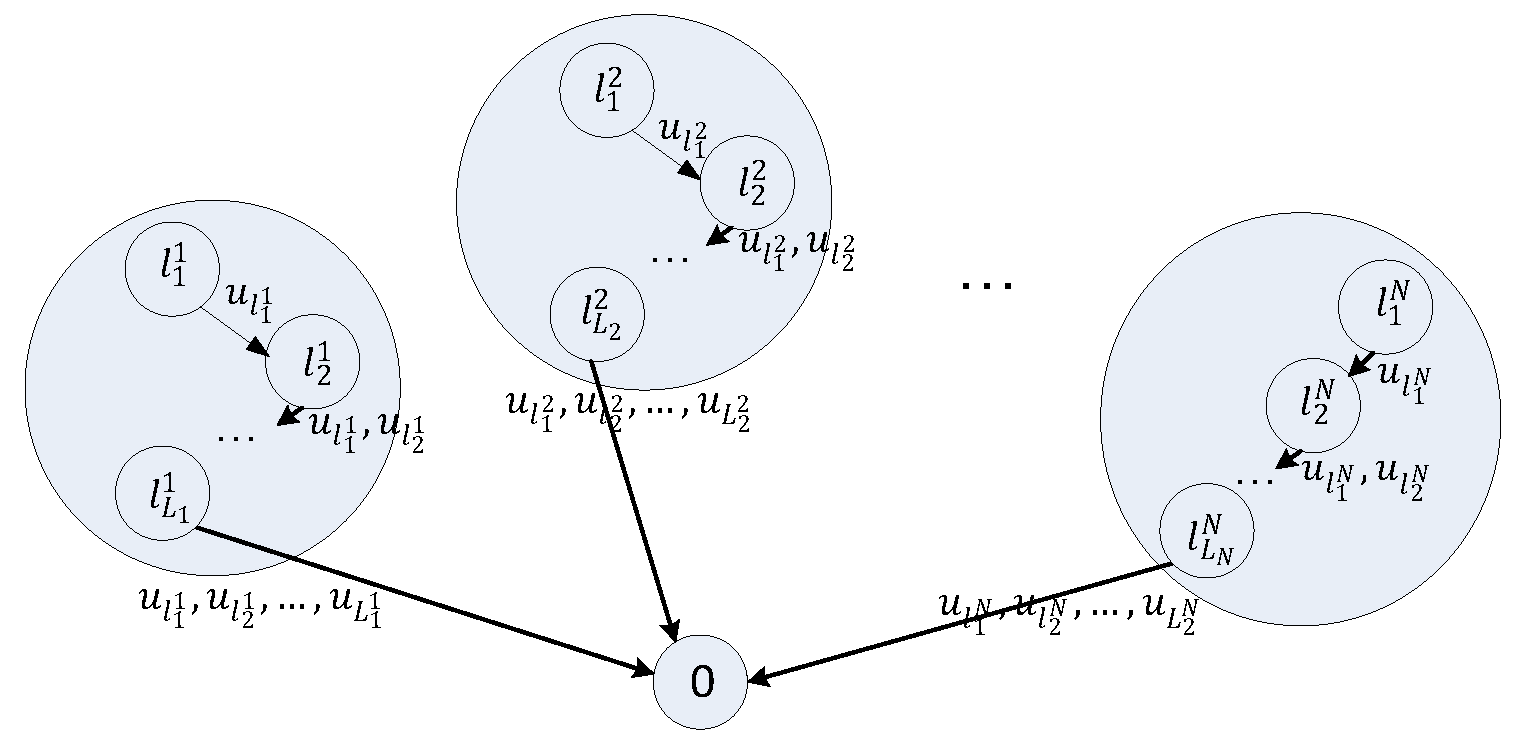}
\caption{Multi-hop Configuration of Decentralized Detection}
 \label{fig:hierar}
\end{figure}
\begin{figure}[!t]
\centering
\includegraphics[scale=0.9]{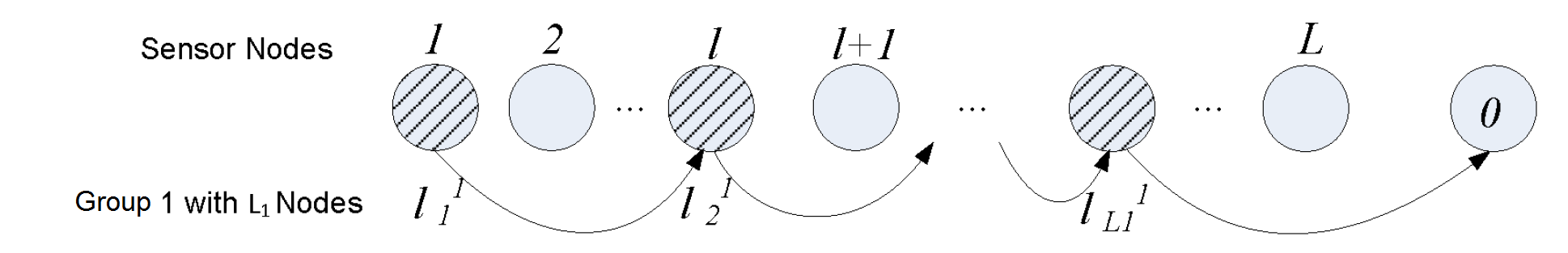}
\caption{$L$ Sensor Nodes Deployed on a Straight Line with the first Group of $L_1$ Nodes}
 \label{fig:grouphier}
\end{figure}
\begin{figure}[!t]
\centering
\includegraphics[scale=0.4]{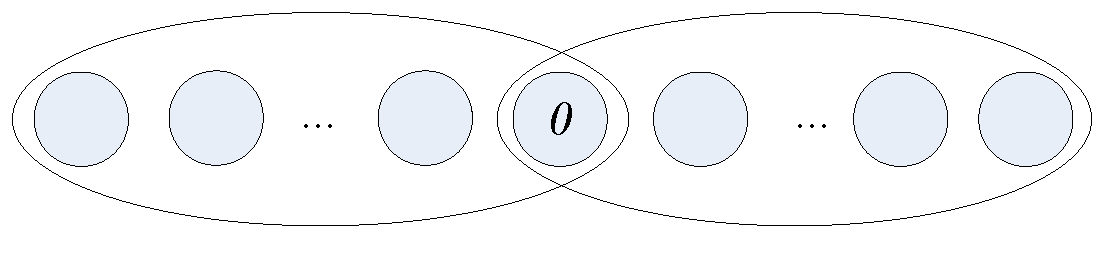}
\caption{Sensor Network with the Fusion Node Located Somewhere Between the Nodes is Divided to Two Sensor Networks.}
 \label{fig:fusionMidd}
\end{figure}

Fig. \ref{fig:hierar} shows the multi-hop configuration of sensor nodes. As shown in the figure, each sensor node sends its quantization bits to the fusion node via multiple hops, which are selected among the other sensor nodes in the network. Fig. \ref{fig:hierar} consists of $N$ groups such that group $n$ consists of $L_n$ sensor nodes. Each sensor node is assigned to a group based on its location in the network. The sum of the sensor nodes in all of the groups is the network size, $L$; in other words we have the equation: $\sum_{n=1}^N L_n=L$. Consider group $n=1,2,...,N$. The first sensor node in this group, $l_1^n$, quantizes its observations and sends the quantization bits to the second sensor node, $l_2^n$. The second sensor node also quantizes its own observations and sends its quantization bits as well as the first node's bits to the third sensor node. This process continues until the last sensor node in the group, $l_{L_n}^n$, quantizes its own observations and sends the quantization bits from the entire sensor nodes in the group, $l_1^n,l_2^n,...,l_{L_n}^n$, to the fusion node. In this configuration, if a sensor node runs out of energy, all of the quantization bits it relays to the fusion node are missed, which results in decreased information in the fusion node. Therefore, it is required that all of the sensor nodes consume the same amount of energy in order to maximize the network lifetime, the time until the first sensor node in the network runs out of energy (section \ref{sec:lifetime}). In addition, in order to maximize information in the fusion node, the energy spent by each sensor node's data to reach the fusion node should be proportional to its contribution to the information in the fusion node (section \ref{sec:infpar}). We solve these two problems with the two degrees of freedom provided by multi-hop configuration: path selection and bit allocation under the constraint of total consumed energy. We propose a method in which best paths are selected to maximize the network lifetime and optimal bits are allocated to maximize the information in the fusion node.\\
Consider a random arrangement of $L$ sensor nodes along with the fusion node on a straight line. Without loss of generality, we assume that the fusion node is located at the end of the line (Fig. \ref{fig:grouphier}). If the fusion node is located in any other position (Fig. \ref{fig:fusionMidd}), we divide the sensor network into two one dimensional networks such that in each of them the fusion node is located at the end of the line. Then the proposed method will be applied on each of them individually. The sensor nodes are ordered from the node farthest from the fusion node to the nearest: $l=1,2,...,L$. In the beginning, the first group is formed. In Fig. \ref{fig:grouphier}, the sensor nodes in the first group are shown by shaded circles. In order to maximize the network lifetime, the following set of equations should be satisfied: ($d_{ij}$ is the distance between the sensor node $i$ and $j$ and $d_i$ is the distance of sensor node $i$ from the fusion node.)
\begin{equation}
\begin{split}
M_{l_1^1} \times d_{l_1^1 l_2^1}^2
=(M_{l_1^1}+M_{l_2^1}) \times d_{l_2^1 l_3^1}^2
=...
= & (\sum_{i=1}^{L_1} M_{l_i^1}) \times d_{l_{L_1}^1}^2\\
= & E/L, \label{eq:1stset}
\end{split}
\end{equation}
in which $d_{l_{L_1}^1}$ is the distance of the last sensor node in the first group from the fusion node.\\
According to the explanation given in section \ref{sec:infpar} for maximizing the information in the fusion node, the energy required for $M_{l_i^1}, i=1,2,...,L_1$ to reach the fusion node, should be proportional to $C_{l_i^1}(M_{l_i^1})$. The consumed energy for transmitting $M_{l_i^1}$ through multiple hops in the group is equal to $M_{l_i^1} \times (\sum_{j=i}^{(L_1-1)}{d_{l_j^1 l_{j+1}^1}^2} + d_{l_{L_1}^1}^2)$. Therefore, we have the following set of equations for the entire sensor nodes in the group: 
\begin{equation}
\begin{split}
& \frac{M_{l_1^1} \times (\sum_{j=1}^{(L_1-1)}{d_{l_j^1 l_{j+1}^1}^2} + d_{l_{L_1}^1}^2)}{C_{l_1^1}(M_{l_1^1})}\\
= & \frac{M_{l_2^1} \times (\sum_{j=2}^{(L_1-1)}{d_{l_j^1 l_{j+1}^1}^2} + d_{l_{L_1}^1}^2)}{C_{l_2^1}(M_{l_2^1})}\\
= & ...\\
= & \frac{M_{l_i^1} \times (\sum_{j=i}^{(L_1-1)}{d_{l_j^1 l_{j+1}^1}^2} + d_{l_{L_1}^1}^2)}{C_{l_i^1}(M_{l_i^1})}\\
= & ...\\
= & \frac{M_{l_{L_1}^1} d_{l_{L_1}^1}^2}{C_{l_{L_1}^1}(M_{l_{L_1}^1})}. \label{eq:2ndset}
\end{split}
\end{equation}
There are $2L_1-1$ independent equations in the above two equation sets \ref{eq:1stset} and \ref{eq:2ndset}. If the first node of the group ($l_1^1$) and its allocated bits ($M_{l_1^1}$) are predefined, then the rest of the nodes in the group (determined by $d_{l_1^1 l_2^1},d_{l_2^1 l_3^1},...,d_{l_{L_1-1}^1 l_{L_1}^1}$ and $d_{l_{L_1}^1}$), and their allocated bits ($M_{l_2^1},M_{l_3^1},...,M_{l_{L_1}^1}$) are $2L_1-1$ unknowns that can be solved by the $2L_1-1$ equations. However, explicit solutions cannot be found by solving this system of equations for a one dimensional network with sensor nodes randomly located on a straight line. Algorithm \ref{alg:hier} is proposed for assigning the sensor nodes to the first group and determining their allocated bits. This algorithm uses equation set \ref{eq:1stset} to assign nodes to the group and equation set \ref{eq:2ndset} to determine the bit allocation. For initialization, the farthest node from the fusion node is chosen as the first node of the first group with one bit allocated to it: $l_1^1=1, M_{l_1^1}=1$. At each step, the next hop for the current sensor node of the group is selected among the remaining sensor nodes in the network and then the new hop's bit allocation is determined. For instance, assume step $i$ with the current sensor node of $l_i^1$. First, the sensor node $l_{i+1}^1$ is defined by calculating $d_{l_{i}^1 l_{i+1}^1}$, the distance between sensor node $l_i^1$ and $l_{i+1}^1$ using equation set \ref{eq:1stset}. Then, $M_{l_{i+1}^1}$ is determined using equation set \ref{eq:2ndset}. At step $i$, the remaining sensor nodes of the group, $l_j^1, j=i+2,i+3,...,L_1$ are not yet defined, so $d_{l_j^1 l_{j+1}^1}, j=i+1,i+2,...,L_1-1$ and also $d_{l_{L_1}^1}$ are still unknown. Therefore, in using equation \ref{eq:2ndset}, the distance that the output of sensor node $l_{i+1}^1$ (which contains all of the quantization bits of $M_{l_j^1}, j=1,2,...,i+1$) travels to reach the fusion node, is approximated by $d_{l_{i+1}^1}$. This process continues until the distance obtained from equation set \ref{eq:1stset} is bigger than the distance between the current sensor node and the fusion node. After forming the first group, the second group is formed similarly by considering the original sensor network excluding the sensor nodes in the first group. The next groups are formed in the similar way. The procedure ends when the entire sensor nodes are grouped.\\
The analysis for Neyman-Pearson formulation is similar to the analysis given for Bayesian formulation. The only difference is considering the Kullback-Leibler divergence instead of Chernoff Information. With the assumption of conditionally independently distributed observations, we have:
\begin{equation}
D(p(\textbf{u}|H_0) || p(\textbf{u}|H_1))=\sum_{l=1}^L(\sum_{u_l=1}^{2^{M_l}}( p(u_l|H_0) \log(\frac{p(u_l|H_0)} {p(u_l|H_1)}))),
\end{equation}
from which the contribution of each sensor and its upper bound are:
\begin{eqnarray}
\lefteqn{ \sum_{u_l=1}^{2^{M_l}}( p(u_l|H_0) \log(\frac{p(u_l|H_0)} {p(u_l|H_1)})) \leq {} }
\nonumber\\
& & {}\int ( f(y_l|H_0) \log(\frac{f(y_l|H_0)} {f(y_l|H_1)})) dy_l.{}
\end{eqnarray}
\begin{algorithm}[!t]
\caption{Determining the first group of multi-hop configuration and their allocated bits for a one dimensional sensor network}
\label{alg:hier}
  \begin{algorithmic}[1]
	\State $i \gets 1$
              \State $l_1^1 \gets 1, M_{l_1^1} \gets 1$
	\While{$l_i^1 < L$}
	\State current sensor node $ \gets l_i^1$
	\State $d_{l_i^1 l_{i+1}^1} \gets \sqrt[2]{\frac{E/L}{\sum_{j=1}^i M_{l_j^1}}}$ \label{alg:hehe}
	\If{$d_{l_i^1 l_{i+1}^1} < d_{l_i^1}$}
	\State Find a sensor node located between the sensor node $l_i^1$ and the fusion node as the next sensor node, $l_{i+1}^1$. This node should be the farthest sensor node from the current sensor node whose distance from the current sensor node is less than or equal to $d_{l_i^1 l_{i+1}^1}$.
	\If{No sensor node found}
	\State $j=\operatorname*{arg\,max}_k (M_{l_k^1}), k=1,2,...,i$
	\State $M_{l_j^1} \gets M_{l_j^1}-1$
	\If{$\sum_{j=1}^i M_{l_j^1} \equiv 0$}
	\State Set the nearest sensor node to sensor node $l_i^1$, which is located between sensor node $l_i^1$ and the fusion node as $l_{i+1}^1$.
	\State $M_{l_{i+1}^1} \gets 1$.
	\State $i \gets i+1$.
	\EndIf
	\State Go to step \ref{alg:hehe}
	\Else
	\State $M_{l_{i+1}^1} \gets \operatorname*{arg\,min}_{M_{l_{i+1}^1}} ((\frac{d_{l_{i+1}^1}^2}{d_{l_i^1 l_{i+1}^1}^2+d_{l_{i+1}^1}^2} \times {C_{l_i^1}(M_{l_i^1})/M_{l_i^1}} ) - C_{l_{i+1}^1}(M_{l_{i+1}^1})/M_{l_{i+1}^1})$.
               \State $i \gets i+1$.
	\EndIf
	\Else
	\If{$i \equiv 1$}
	\State $M_{l_{i}^1} \gets \left\lfloor \frac{E/L}{d_{l_1^1}^2} \right\rfloor$
	\EndIf
	\State Go to \ref{alg:endd}
	\EndIf
	\EndWhile
	\State The first group with the size of $i$ nodes and their allocated bits are defined. \label{alg:endd}
  \end{algorithmic}
\end{algorithm}
Using the procedure similar to what is discussed for Bayesian formulation, the optimum decision regions at each sensor node are obtained from:
\begin{equation}
\gamma_l^{\text{opt}}(M_l)=\operatorname*{arg\,max}_{\gamma_l(M_l)} \sum_{u_l=1}^{2^{M_l}}( p(u_l|H_0) \log(\frac{p(u_l|H_0)} {p(u_l|H_1)})).
\end{equation}
Therefore, in Neyman-Pearson problem formulation, $\sum_{u_l=1}^{2^{M_l}}(p(u_l|H_0) \log(\frac{p(u_l|H_0)} {p(u_l|H_1)}))$ should be used instead of $C_l(M_l)$.
\subsection{Comparison Between Parallel and Multi-hop Configurations}
Direct transmission consumes more energy than transmission via multiple hops. Therefore, the total consumed energy in multi-hop configuration is less than parallel configuration with the same bit allocation among the sensor nodes. Equivalently, under the constraint of total consumed energy, more bits can be allocated to the sensor nodes in multi-hop configuration than the parallel configuration, which results in more information in the fusion node. In addition, as Chernoff information and Kullback-Leibler divergence are concave functions of number of bits, the information obtained with the first allocated bit is high (more than 60\% of the upper bound as shown in Fig. \ref{fig:Gauss}) and with more number of bits, the increase in the information decreases. Therefore, for a fixed number of allocated bits, even distribution of bits among the sensor nodes results in more information than uneven distribution. In parallel configuration, more bits are allocated to the closer nodes to the fusion node and fewer bits are allocated to the farther nodes. While in proposed multi-hop configuration, bits are allocated among the sensor nodes more evenly. Thus, for a fixed number of allocated bits, the performance of multi-hop configuration is better than parallel configuration in terms of information in the fusion node (as shown in Fig. \ref{fig:compare2}).\\
We briefly examine the delay incurred in multi-hop and parallel configurations. We assume that all sensor nodes use identical wireless transmission technologies. In wireless communication over a parallel configuration for a one dimensional sensor network with the fusion node located at one end, the sensor nodes cannot send their data simultaneously in order to avoid collisions when their transmission ranges overlap. As a result the maximum delay is $O(L)$, where $L$ is the number of nodes. In our proposed multi-hop configuration, the transmissions are performed through multiple hops. As a result the maximum delay is at worst equal to that of the parallel configuration.\\
In the case of a link failure from sensor node $l_1$ to sensor node $l_2$ in multi-hop configuration, the entire information that sensor node $l_1$ should send to the fusion node is missed in the fusion node. If sensor node $l_1$ plays the role of intermediate node for other nodes, then the information from all of those nodes cannot reach the fusion node. While, in parallel configuration, the information of just one sensor node is missed in the fusion node in the case of link failure between any sensor node and the fusion node. From this point of view, multi-hop configuration is more sensitive to link failure than the parallel configuration.\\
\section{Simulation Results} \label{sec:five}
The performance of the proposed multi-hop configuration for the decentralized detection problem is evaluated by simulating in MATLAB and compared with the parallel configuration in terms of energy cost and information quality. Two sets of results are provided for the case of parallel configuration, one based on maximizing chernoff information and another based on maximizing network lifetime. We simulated Gaussian random variables of observations in the sensors; $f(y|H_0) \sim \mathcal{N}(-1,1)$ and $f(y|H_1) \sim \mathcal{N}(1,1)$ are considered for determining optimized decision rules within the sensor nodes. We assume a single dimensional WSN field, where the nodes are randomly deployed on a straight line. The fusion node is located within $2$ units distance from the end point of the line. All the results are average values over 1000 iterations.
\subsection{Effect of Quantization} \label{sec:Gausss}
\begin{figure}[!t]
\centering
\includegraphics[scale=0.25]{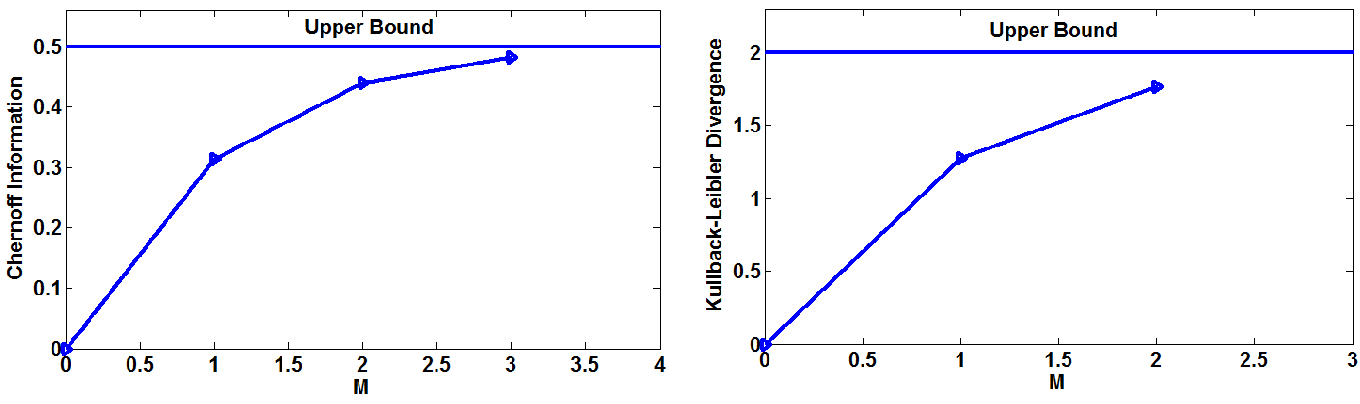}
\caption{Optimized (a) Chernoff Information, and (b) Kullback-Leibler Divergence, for Different Values of Allocated Bits}
 \label{fig:Gauss}
\end{figure}

First, we need to determine decision rules of all the nodes for different values of $M$ (number of allocated bits). Based on the fact that Gaussian distributions satisfy monotone likelihood ratio property, we determine quantization thresholds at each sensor node accordingly, for different values of $M$. Since the probability distributions at all of the nodes are identical, the optimization problem for all of the nodes is identical and is expressed as:
\begin{equation}
C_{opt}(M)=\max_{t_1,t_2,...t_{2^M-1}}{C_l(M)},
\end{equation}
and
\begin{equation}
\max_{t_1,t_2,...t_{2^M-1}}{\sum_{u_l=1}^{2^{M_l}}( p(u_l|H_0) \log(\frac{p(u_l|H_0)}
{p(u_l|H_1)}))}
\end{equation}
for Bayesian and Neyman-Pearson formulations respectively. Here $t_i, i=1,2,...,2^M-1$ are the thresholds that have the property $t_1<t_2<...<t_{2^M-1}$ under the monotone likelihood ratio. The optimum thresholds for different number of allocated bits were obtained by simulation and are listed in Tables \ref{table:nonlin} and \ref{table:neyman} with four-digit accuracy of Chernoff information and Kullback-Leibler divergence. As the simulation results show, Fig. \ref{fig:Gauss}, the Chernoff information and Kullback-Leibler divergence are increasing concave functions of the number of allocated bits with the upper bounds of $0.5$ and $2.0$, respectively.
\subsection{Multi-hop vs Parallel Configuration}
\begin{table}[!t]
\caption{Optimized Decision Rules at Each Sensor Node for Different Values of $M$,
Gaussian Distribution of Bayesian Formulation} 
\centering  
\begin{tabular}{c c c} 
\hline\hline                        
$M$ & thresholds & $C_{opt}(M)$\\ [0.5ex] 
\hline                  
1 & [0] & 0.3137  \\ 
2 & [-1 0 1] & 0.4399  \\
3 & [-1.8        -1.1 -0.5 0        0.5        1.1         1.8] & 0.4824  \\[1ex]      
\hline 
\end{tabular}
\label{table:nonlin} 
\end{table}
\begin{table}[!t]
\caption{Optimized Decision Rules at Each Sensor Node for $M=1,2$, Gaussian
Distribution of Neyman-Pearson Formulation} 
\centering  
\begin{tabular}{c c c} 
\hline\hline                        
$M$ & thresholds & $\sum_{u_l=1}^{2^{M_l}}( p(u_l|H_0) \log(\frac{p(u_l|H_0)}
{p(u_l|H_1)}))$\\ [0.5ex] 
\hline                  
1 & [-0.6] & 1.2788  \\ 
2 & [-1.7 -0.7 0.3] & 1.7653  \\[1ex]      
\hline 
\end{tabular}
\label{table:neyman} 
\end{table}
\begin{figure}[!t]
\centering
\includegraphics[scale=0.35]{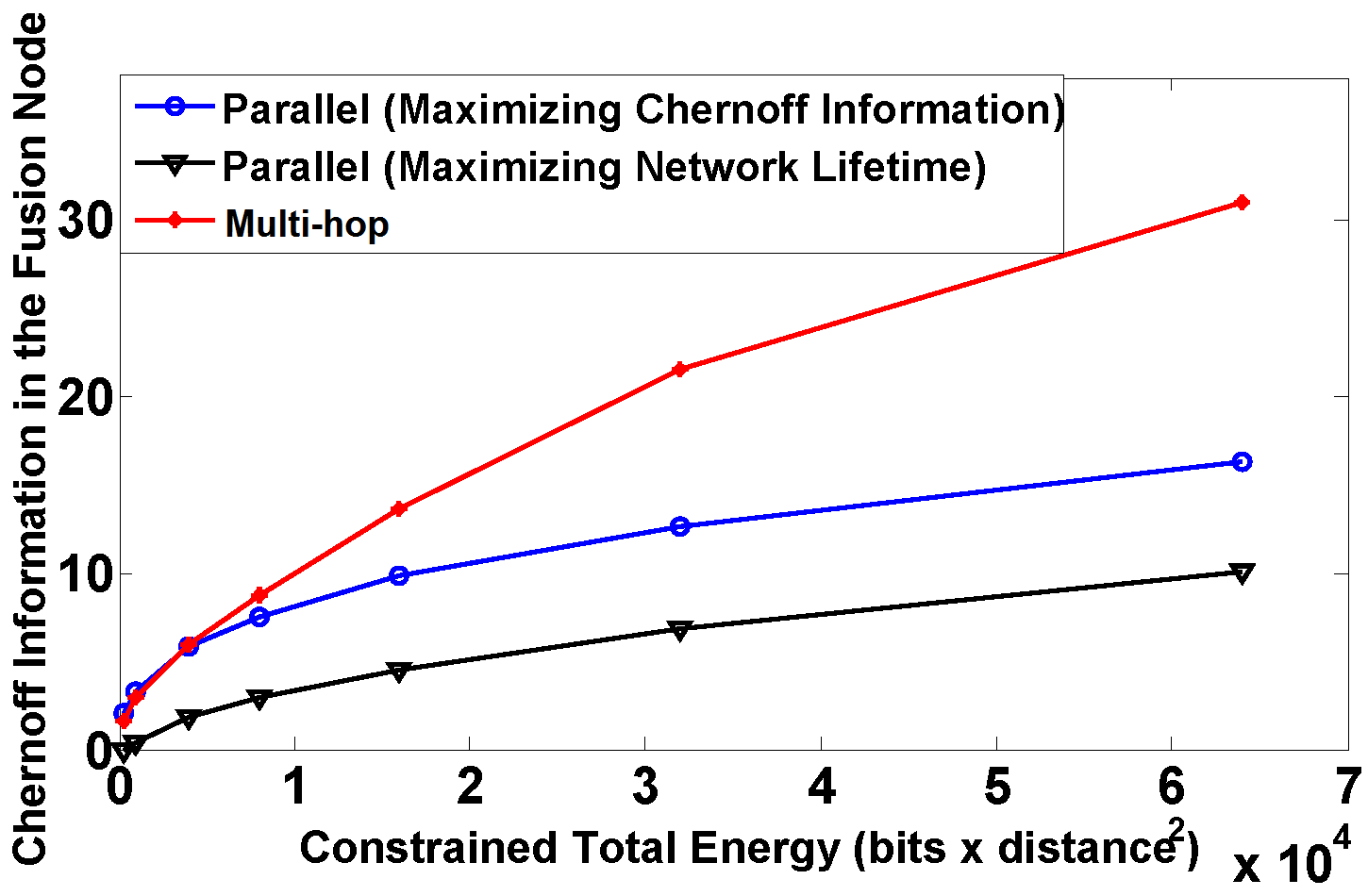}
\caption{Chernoff Information in the Fusion Node Versus Constrained Total Energy}
 \label{fig:compare1}
\end{figure}
\begin{figure}[!t]
\centering
\includegraphics[scale=0.37]{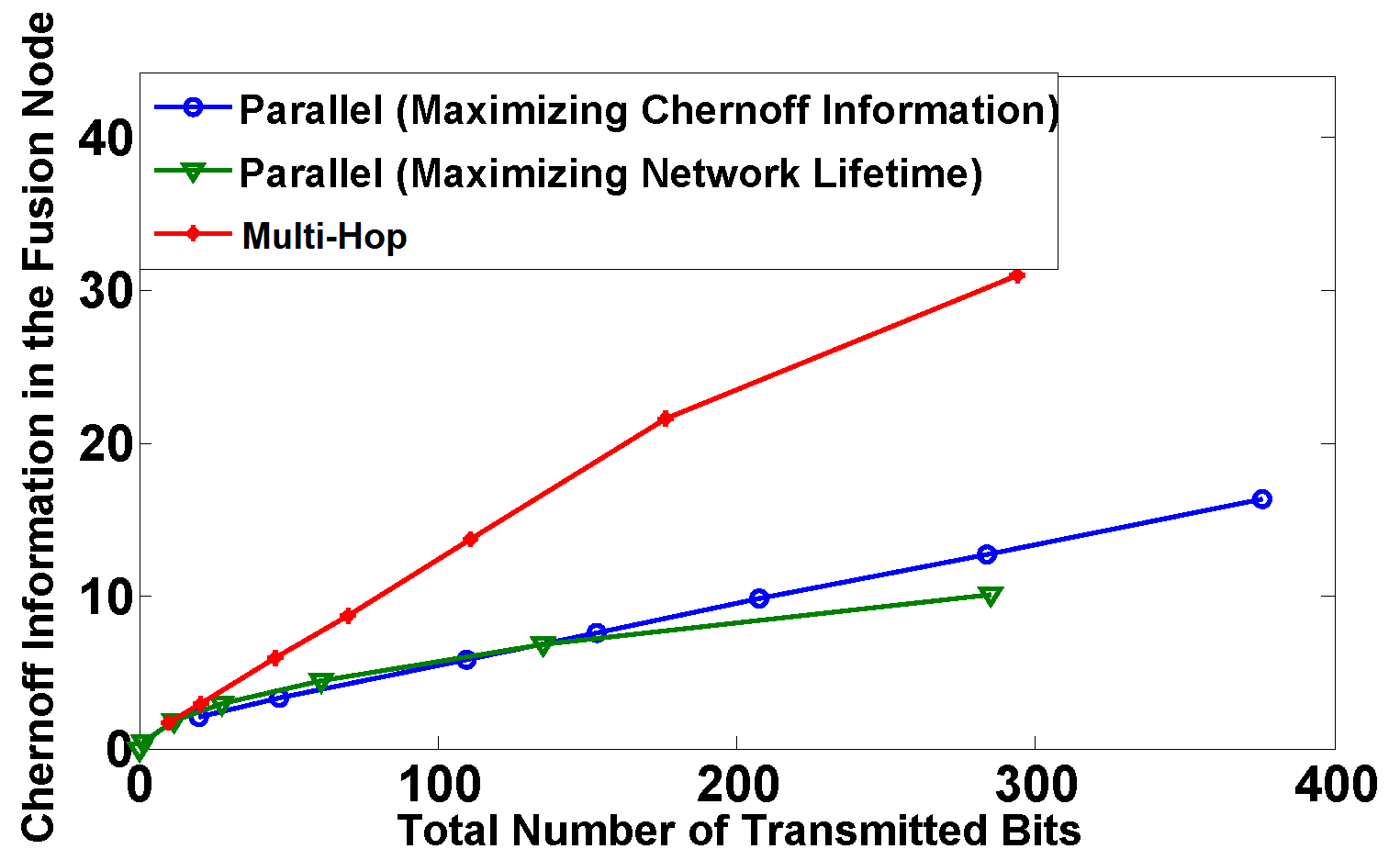}
\caption{Chernoff Information in the Fusion Node Versus Total Transmitted Bits}
 \label{fig:compare2}
\end{figure}
In this section, we consider $100$ sensor nodes deployed uniformly with the density of one node per unit length. The results obtained in the previous section are used in this section to simulate the performances of the proposed methods for parallel and multi-hop configurations. We have simulated our proposed methods for both Bayesian and Neyman-Pearson formulations and the results show very similar trends for both cases. In this paper, we report only data for Bayesian formulation. Fig. \ref{fig:compare1} shows the amount of Chernoff Information in the fusion node for different values of the constrained total energy in parallel and multi-hop configurations. As shown in the figure, the Chernoff Information in the fusion node obtained from the multi-hop configuration is more than the Chernoff Information obtained from the parallel configuration. Significant improvement is obtained for larger constrained total energy. Fig. \ref{fig:compare2} shows the Chernoff Information versus the total number of transmitted bits to the fusion node for multi-hop and parallel configurations. These curves are obtained by changing the value of constrained total energy and using the proposed methods for determining the bit allocation amongst nodes and computing the related Chernoff information in the fusion node. The allocated bits are divided amongst the sensor nodes more evenly in the multi-hop configuration than the parallel configuration, resulting in more Chernoff Information in the fusion node. Higher improvements were shown by considering larger total number of transmitted bits.
\subsection{Effect of Scaling the Network Size}
\begin{figure}[!t]
\centering
\includegraphics[scale=0.45]{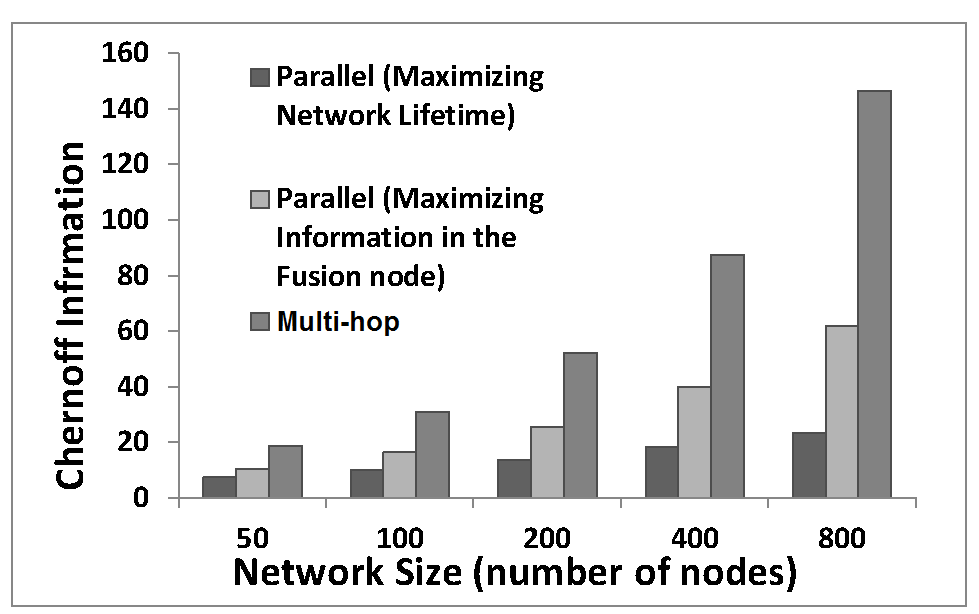}
\caption{Chernoff Information in the Fusion Node for Different Network Sizes}
 \label{fig:density}
\end{figure}

In this section, we investigate the effect of network size on the performance improvement in the multi-hop configuration compared with the parallel configurations. Simulations are performed with the constrained total energy of $E=64000 (bits \times distance^2)$. Fig. \ref{fig:density} compares performances of the multi-hop and the parallel configurations for different network sizes. The results are presented for $L=50,100,200,400,800$ sensor nodes, which are uniformly deployed on a straight line with the length of $100$ units. We see that larger networks show significantly more improvement in terms of Chernoff Information in the fusion node.
\section{Conclusion} \label{sec:concl}
In this paper, an energy efficient decentralized detection was studied using a multi-hop configuration of the sensor nodes. We formulated the problem to achieve two objectives: maximizing information in the fusion node and maximizing network lifetime. We showed that in parallel configuration, where each node sends its data directly to the fusion node, the stated objectives cannot be simultaneously obtained. Whereas, in multi-hop configuration, these two objectives were achieved simultaneously using multi-hop transmission of data. Under the constraint of total energy, optimal bit allocations amongst the sensor nodes were proposed for parallel and multi-hop configurations. By taking advantage of optimal bit allocation amongst the sensor nodes, considerable improvements in terms of information quality and energy efficiency were achieved in the fusion node in multi-hop configuration as compared with parallel configuration.
 \bibliographystyle{IEEEtran}
 \bibliography{test}

\end{document}